\documentstyle[preprint,epsfig,aps]{revtex} 
\draft
\begin{document}\tighten
\title{Correlated adiabatic and isocurvature perturbations from double 
inflation}
\author{David Langlois}
\address{D\'epartement d'Astrophysique Relativiste et de Cosmologie (UPR 176),\\
Centre National de la Recherche Scientifique,\\
Observatoire de Paris, 92195 Meudon Cedex, France}
\date{\today} 
\maketitle

\def\k{{\bf k}}
\def\x{{\bf x}}
\def\e{{\bf e}}
\def\beq{\begin{equation}}
\def\eeq{\end{equation}}
\def\d{{\delta}}
\def\dc{\d^{(c)}}
\def\noneq{\neq}

\par\bigskip

\begin{abstract}
It is shown that double inflation (two minimally coupled massive 
scalar fields) can produce {\it correlated adiabatic 
and isocurvature primordial perturbations}. Depending on the two relevant 
parameters  of the model, the contributions to the primordial 
perturbations are computed, with special emphasis on the correlation, which 
can be quantitatively represented by a correlation spectrum.  Finally 
the primordial spectra are evolved numerically to obtain the CMBR
anisotropy multipole expectation values. It turns out  that the existence 
of mixing and correlation can alter very significantly the temperature 
fluctuation predictions. 
\end{abstract}

\section{Introduction}

In our present picture,  cosmological fluctuations today are seen as the 
 {\it combination} of an initial spectrum, which  can be computed within the 
framework of high energy models (like inflation),  
 with the subsequent  processes occuring at lower energy  where the 
physics is believed to be better understood (up to some unknowns such as the amount 
and nature of dark matter, the mass of neutrinos...). In the near future, we expect 
new and {\it precise} information on cosmological fluctuations with the planned measurements
of the Cosmic Microwave Background Radiation (CMBR) anisotropies by the 
satellites MAP \cite{map} and PLANCK \cite{planck}. It has been emphasized in the 
recent years that the precision of these measurements could in principle allow 
us to determine with a high precision the cosmological parameters 
\cite{parameters}. These studies 
however all assume very simple initial perturbations, typically 
Gaussian adiabatic perturbations with a power-law spectrum. 
However, reality could turn out to be more subtle.
This then would have the drawback  
to complicate the determination of the cosmological parameters but could 
open the fascinating perspective  to gain  precious information 
on the primordial universe.  At present, at a time when data are still 
unprecise, it is essential   to identify broad categories of early universe
models and to determine their specificities as far as observable quantities
are concerned, with the purpose to be able to discriminate between these
various classes of models when detailed data will become available.

 Ultimately, inflation must be related to a high energy physics model.
Today there are many viable models but  a generic feature of these
models is that they contain generally many scalar fields. A property 
of inflation with several scalar fields is that it can generate, in 
addition to the ubiquitous adiabatic perturbations, isocurvature 
perturbations. In this respect, it is important to consider the possible role 
of primordial  isocurvature perturbations. Isocurvature perturbations are perturbations
in the relative density ratio between various species in the early universe, in 
contrast with the more standard adiabatic (or isentropic) perturbations which are 
perturbations in the total energy density with fixed particle number ratios.
Primordial isocurvature perturbations are, most of the time, ignored in 
inflationary models. The main reason for this is that they are less universal
than adiabatic perturbations because, on one hand, they can be produced only 
in multiple inflationary models \cite{multinf}, 
and, on the other hand, they can survive 
until the present epoch only if at least one of the inflaton fields remains
decoupled from ordinary matter during the whole history of the universe. 
However, not only the existence of isocurvature perturbations is allowed in 
principle, but   candidates for  inflatons with the required above conditions even 
exist in many theoretical models (dilatons, axions).

What has already been established is that  a  pure isocurvature scale-invariant 
spectrum must be rejected because it predicts on large scales 
too  large temperature anisotropies with respect to density fluctuations \cite{eb86}. 
But other possibilities can be envisaged. Several have been investigated 
in the literature: tilted isocurvature perturbations \cite{tilted-iso}, 
combination of isocurvature and adiabatic perturbations \cite{mixing}.
In the latter case, only combinations of {\it independent} isocurvature 
and adiabatic perturbations were considered. The aim of this paper  is 
to investigate the possibility of {\it correlated mixtures of isocurvature 
and adiabatic} perturbations.

To illustrate this, the simplest model of multi-field 
inflation is considered here: double inflation \cite{dblinf}, namely a model with two
massive scalar fields without self-interaction or mutual interaction 
(other than gravitational). The production of fluctuations in this model has
been studied in great detail by Polarski and Starobinsky and, in the present work, 
their notation and formalism will be followed closely.  They were interested 
essentially in  adiabatic perturbations \cite{ps92} (see also \cite{p94} for 
a numerical analysis)
but also considered isocurvature perturbations \cite{ps94}. 
However, they did not investigate the range of parameters where this model has the 
striking property to produce correlated isocurvature and adiabatic perturbations. By 
this, we mean the cases where both isocurvature and adiabatic perturbations receive 
significant contributions of at least one of the scalar fields, in contrast to the 
uncorrelated case where one of the scalar fields feeds essentially the adiabatic 
perturbations while the second one is at the origin of the isocurvature perturbations.

The plan of this paper is the following. In section 2, the model 
of double inflation will be presented. Section 3 will be devoted to the 
analysis of adiabatic and isocurvature perturbations: their definition, 
how they are obtained from the inflation perturbations, the conditions 
to obtain correlated mixtures. Section 4 considers formally the definition 
of spectra for the perturbations as well as the notion of correlation. 
In section 5, the predictions for the CMBR anisotropies and matter
power spectrum are given for the models with correlated primordial 
perturbations.

\section{Double inflation}

As mentioned in the introduction, inflation needs at least two scalar
fields to produce isocurvature perturbations. That is why we investigate
the simplest model of inflation with two scalar fields: they are
non-interacting, massive, minimally coupled scalar fields. The Lagrangian
corresponding to this model is
\begin{equation}
{\cal L}={^{(4)}R\over 16\pi G}-{1\over 2}\partial_{\mu}\phi_l\partial^{\mu}
\phi_l-{1\over 2}m^2_l\phi^2_l - {1\over 2}\partial_{\mu}\phi_h\partial^{\mu}
\phi_h - {1\over 2}m_h^2 \phi^2_h  ,
\end{equation}
where the subscripts $l$ and $h$ designate respectively the light and heavy
scalar fields (and thus $m_h>m_l$). $^{(4)}R$ is the scalar spacetime
curvature and G Newton's constant.

\subsection{The background equations}

In a spatially flat Friedmann-Lema\^ itre-Robertson-Walker (FLRW) spacetime, 
with metric $ds^2=-dt^2+a^2(t)d\vec{x}^2$,
the equations of motion read
\begin{eqnarray}
3H^2&=&4\pi G(\dot{\phi}_l^2+\dot{\phi}_m^2+m_l^2\phi^2_l+m_h^2\phi^2_h)
   \label{2a}\\
\ddot{\phi}_l&+&3H\dot{\phi}_l+m^2_l\phi_l=0, \nonumber  \\
\ddot{\phi}_h&+&3H\dot{\phi}_h+m^2_h\phi_h=0. \label{eom}
\end{eqnarray}
Following \cite{ps92} it is convenient, during the phase when both
scalar fields are slow-rolling (i.e. when $\dot{\phi}_l^2$ and $\dot{\phi}_h^2$
can be neglected in (\ref{2a}), $\ddot{\phi}_l$ and $\ddot{\phi}_h$ in 
(\ref{eom})),
to write the evolution of the two scalar fields in  the following  
parametric form 
\begin{equation}
\phi_h=\sqrt{s\over 2\pi G} \sin \theta ,\qquad  \phi_l=\sqrt{s\over 2\pi G}
\cos \theta
\label{3}
\end{equation}
where 
\beq
s=-\ln (a/a_e)
\eeq
 is the number of e-folds between a given instant 
and the end of inflation. This form  (\ref{3}) is a consequence of the 
approximate relation $d(\phi_h^2+\phi_l^2)/ds=-d(\phi_h^2+\phi_l^2)/(Hdt)
\simeq (2\pi G)^{-1}$ resulting from the (slow-roll) equations of motion. 
The angular variable $\theta$ 
can then be related to the parameter $s$ by the expression
\begin{equation}
s=s_0{(\sin \theta)^{2\over R^2-1}\over (\cos \theta)^{2R^2\over R^2-1}}
\label{stheta}
\end{equation}
where $R$ is the ratio of the masses of the two scalar fields
\beq
R\equiv {m_h\over m_l}.
\label{5}
\eeq
The equation (\ref{stheta}) was obtained by integrating the relation 
giving $d\theta/d(\ln s)$ as a function of $\theta$, which can be established 
by use of the slow-roll approximation of the equations of motion (\ref{2a})-(\ref{eom})
(see \cite{ps92} for the details of the calculations). As noticed 
in \cite{ps92}, equation (\ref{stheta}) can be approximated, when 
$\theta\gg R^{-1}$, by the simple formula
\beq
s\simeq {s_0\over \cos^2\theta}. \label{5bis}
\eeq
This behaviour corresponds to the period when inflation is dominated 
by the heavy scalar field (this approximation is valid as long as $s>s_0$
and $s-s_0\gg s_0/R^2$). This period ends when $\theta\sim R^{-1}$, 
$s\sim s_0$ and is followed (possibly after a dust-like transition period) by 
a phase of inflation dominated by the light scalar field.

It follows from (\ref{2a})-(\ref{3}) 
that the Hubble parameter can be expressed in the form
\begin{equation}
H^2(s)\simeq {2\over 3}m^2_l s[1+(R^2-1)\sin^2\theta (s)] ,
\label{6}
\end{equation}
where the function $\theta(s)$ is obtained by inverting (\ref{stheta})
($0<\theta<\pi/2$). As inflation proceeds, $s$ decreases and $\theta$ goes
to smaller and smaller values, which implies a decreasing Hubble parameter 
during inflation. 

It will be convenient to define $s_H$ as the number of e-folds before
the end of inflation when the scale corresponding to our Hubble radius 
today crossed out the Hubble radius during inflation.
The value of $s_H$ depends on the temperature after the reheating
(see e.g. \cite{ll} )  but 
roughly $s_H\simeq 60$. To make definite calculations, we shall take 
 throughout this work the  value
\beq
s_H=60.
\eeq
Note  that the class of models considered here 
depends on three free parameters: the two masses $m_l$ and $m_h$, or 
alternatively $m_l$ and $R$, and the parameter $s_0$. In particular, the choice
of this last parameter $s_0$ relatively to $s_H$ will determine the 
specific phase of double inflation, `heavy' dominated, intermediate or 
`light' dominated, during which the perturbations on scales of cosmological 
relevance were produced.

\subsection{Perturbations}
After having determined the evolution of the background quantities, let us 
turn now to the evolution of the linear perturbations. We shall restrict our
analysis 
to the so-called scalar perturbations (in the terminology of Bardeen 
\cite{bardeen}). 
We thus consider a spacetime linearly perturbed
about the flat FLRW spacetime of the previous subsection, 
endowed with the metric
\begin{equation}
ds^2=-(1+2\Phi)dt^2+a^2(t)(1-2\Psi)\delta_{ij}dx^idx^j. 
\end{equation}
Although this metric is not the most general a priori, it turns out that any 
perturbed metric (of the scalar type) can be transformed into a metric 
of this form by a suitable coordinate transformation. 
This choice corresponds to the so-called longitudinal gauge. In addition 
to the geometrical perturbations $\Phi$ and $\Psi$, one must also consider the matter 
perturbations, which will simply be during inflation 
  the perturbations of the scalar
fields, respectively $\delta\phi_h$ and  $\delta\phi_l$, with respect to 
their  homogeneous values.

Before writing down the equations of motion for the perturbations, it 
is convenient to use a Fourier decomposition and to define the 
Fourier modes of any perturbed quantity $f$ by the relation 
\begin{equation}
f_{\k}=\int{d^3\x\over (2\pi)^{3/2}} e^{-i\k.\x} f(\x).
\end{equation}
The equations of motion for the perturbations are derived from the 
perturbed Einstein equations and from the Klein-Gordon equations of 
the scalar fields. They lead to the 
 following four equations (see e.g. \cite{mfb})
\beq
\Phi=\Psi,
\eeq
\begin{equation}
\dot\Phi+H\Phi=4\pi G\left(\dot\phi_h\delta\phi_h
+\dot\phi_l\delta\phi_l\right),
\end{equation}
\begin{equation}
\ddot{\delta\phi_h}+3H\dot{\delta\phi_h}+\left({k^2\over a^2}+m_h^2\right)\delta\phi_h=
4\dot\phi_h\dot\Phi-2m_h^2\phi_h\Phi, \label{eqmperth}
\end{equation}
\begin{equation}
\ddot{\delta\phi_l}+3H\dot{\delta\phi_l}+\left({k^2\over a^2}+m_l^2\right)
\delta\phi_l=
4\dot\phi_l\dot\Phi-2m_l^2\phi_l\Phi, \label{eqmpertl}
\end{equation}
where the subscript $\k$ is here implicit, as it will be throughout this paper.

In the slow-rolling approximation and for superhorizon modes, i.e.
$k\ll aH$, these equations can be solved (see \cite{ps92}) and the dominant
solutions read
\begin{equation}
\Phi\simeq -{C_1\dot H\over H^2}+2C_3{(m_h^2-m_l^2)m_h^2\phi_h^2m_l^2\phi_l^2\over
3(m_h^2\phi_h^2+m_l^2\phi_l^2)^2}, \label{14}
\end{equation}
\begin{equation}
{\delta\phi_l\over\dot\phi_l}\simeq {C_1\over H}-2C_3{Hm_h^2\phi_h^2\over
m_h^2\phi_h^2+m_l^2\phi_l^2},\quad
{\delta\phi_h\over\dot\phi_h}\simeq {C_1\over H}+2C_3{Hm_l^2\phi_l^2\over
m_h^2\phi_h^2+m_l^2\phi_l^2},\label{15}
\end{equation}
where $C_1(\k)$ and $C_3(\k)$ are time-independent constants
of integration and  are fixed by the initial
conditions. As usual in inflation, perturbations are assumed to be
initially (i.e. before crossing out the Hubble radius) in their vacuum
quantum state. Perturbations outside the Hubble radius are then obtained
by amplification of the vacuum quantum fluctuations due to the gravitational
interaction. The two scalar fields being independent, one simply
duplicates the results of single scalar field inflation (see e.g. 
\cite{mfb}). Consequently,
$\delta\phi_h$ and $\delta\phi_l$ can be written, for wavelengths
crossing out the Hubble radius, as
\begin{equation}
\delta\phi_h={H_k\over \sqrt{2k^3}}e_h(\k), \quad 
\delta\phi_l={H_k\over \sqrt{2k^3}} e_l(\k), \label{16}
\end{equation}
where $e_h$ and $e_l$ are classical Gaussian random fields with
$\langle e_i(\k)\rangle=0$, $\langle e_i(\k)e_j^*(\k')\rangle=\delta_{ij}
\delta(\k-\k')$, for $i,j=l,h$, and 
$H_k$ is the Hubble parameter when the mode crosses the Hubble radius,
i.e., when $k=2\pi aH$. Neglecting the evolution of the Hubble 
parameter with respect to that of the scale factor, the number of e-folds
$s_k$ corresponding to the instant when the mode of wavenumber $k$ 
crossed out the Hubble radius, is given simply by
\beq
k\simeq k_H e^{s_H-s_k}, \label{17}
\eeq
where $k_H$ is the wavenumber corresponding to the present Hubble scale.
In the present work, our interest will focus on the scales of cosmological 
relevance, typically $k/k_H\simeq 0.1-2000$. This means that the range of 
e-folds that will interest us is $52\lesssim s_k \lesssim 63$.

\section{Primordial perturbations}
The analysis of the solutions for the perturbations during inflation obtained
in the previous section will now enable us to determine the ``initial"
(but post-inflationary) conditions for the perturbations in the radiation 
era taking place after inflation and reheating.

\subsection{Initial conditions in the radiation era}
At some past instant deep in the radiation era, we shall consider four species 
of particles.
Two species will be relativistic: photons and neutrinos; two species 
will be  non-relativistic: baryons and cold dark matter. Their
respective energy density contrasts will be denoted $\d_\gamma$, 
$\d_\nu$, $\d_b$ and $\d_c$ ($\d_A\equiv \d\rho_A/\rho_A$).

At this point, it is useful to define precisely the notion of adiabatic and isocurvature 
perturbations. Isocurvature pertubations are defined by the condition that there is 
no perturbation of the energy density in the total comoving gauge (denoted 
by the subscript $(c)$), i.e.
\beq
\sum_A {\d^{(c)}}\rho_A=0, 
\eeq
but that there are perturbations in the ratios of species particle numbers, i.e.
\beq
\dc(n_A/n_B)\noneq 0 \eeq
in general. By contrast, adiabatic (or isentropic) perturbations are defined by 
the prescription that the particle number ratios between various species is fixed, i.e.
\beq
\dc(n_A/n_B)= 0 ,
\eeq
whereas the total energy density perturbation can fluctuate, i.e. 
\beq
\sum_A {\d^{(c)}}\rho_A\noneq 0.
\eeq
It is clear from the above definitions that, if one considers N species, there will 
be in general one adiabatic mode and $N-1$ isocurvature modes. Here, it will be assumed 
that the light scalar field $\phi_l$ decays into ordinary particles, i.e. gives birth 
to the photons, neutrinos and baryons, while the dark matter particles are associated
exclusively with the heavy scalar field $\phi_h$. Note that 
part of the dark matter could be also 
produced by the light scalar field but we shall ignore this possiblity here
for simplicity. As a consequence, the particle number ratios
between the three `ordinary' species will be frozen, i.e. 
\beq
{\dc n_\gamma\over n_\gamma}= {\dc n_\nu\over n_\nu}={\dc n_b\over n_b},
\eeq
and only one isocurvature mode will exist, which can be conveniently represented by the 
quantity
\beq
S\equiv {\dc n_c\over n_c}-{\dc n_\gamma\over n_\gamma}= \dc_c-{3\over 4}\dc_\gamma.
\label{23}
\eeq

Going back to the longitudinal gauge, and following Ma and Bertschinger \cite{mb95}, one can 
write the initial conditions deep in the radiation era for modes outside the 
Hubble radius in the form 
\begin{eqnarray}
& &\d_\gamma=-2\Phi,\cr
& & \d_b={3\over 4}\d_\nu={3\over 4}\d_\gamma,\cr
& & \d_c=S+{3\over 4}\d_\gamma, \cr
& & \theta_\gamma=\theta_\nu=\theta_b=\theta_c={1\over 2}(k^2\eta)\Phi, \cr
& & \sigma_\nu={1\over 15}(k\eta)^2\Phi, \cr
& & \Psi=\left(1+{2\over 5}R_\nu\right)\Phi,  
\end{eqnarray}
with $R_\nu=\rho_\nu/(\rho_\gamma+\rho_\nu)$ and where $\theta$ stands 
for the divergence of the fluid three-velocity, 
$\sigma_\nu$ for the shear  stress of neutrinos (in the rest of this paper, the contribution of neutrinos in all analytical calculations will be ignored for simplicity but it will be taken into account in the numerical 
calculations) and $\eta$ is the conformal 
time defined by $d\eta=dt/a(t)$. All the information about the initial 
conditions is thus contained in the two k-dependent quantities $\Phi$ and $S$, which are time-independent during the radiation era (for $k\ll aH$).
The next subsections will be devoted to make the link between these two
quantities and 
the  perturbations during inflation.

\subsection{Adiabatic initial perturbations}
The evolution of $\Phi$ is given by (\ref{14}) only during the phase when
both scalar fields are slow rolling. As soon as the heavy scalar field $\phi_h$
ends its slow-rolling phase, the second term on the right hand side of (\ref{14})
will die out, and $\Phi$ can be given during all the subsequent evolution 
of the universe in the simple form 
\beq
\Phi=C_1\left(1-{H\over a}\int_0^t a(t') dt'\right). \label{25a}
\eeq
It can be checked that, during inflation, $1-{H\over a}\int_0^t a(t') dt'
\simeq -\dot H/H^2$, which ensures that the coefficient $C_1$ in the 
above formula is the same as in (\ref{14}).

It is thus essential to express $C_1$ in terms of the perturbations of the 
two scalar fields, and therefore in terms of $e_l$ and $e_h$, in order 
to be able 
to determine the amplitude of the perturbations after the end of inflation.
Combining the two equations in (\ref{15}) and using (\ref{16}), as well 
as the slow-roll approximation of the background equations of motion 
(\ref{2a})-(\ref{eom}), the expression for the coefficient $C_1$ during inflation 
is found to be
\beq
C_1(\k)\simeq -{4\pi G}{H_k\over \sqrt{2k^3}}\left[\phi_l e_l(\k)+\phi_h e_h(\k)
\right].
\eeq
As it is clear from this formula, $C_1(k)$ is a stochastic variable, whose
properties can be determined from the stochastic properties 
of $e_l$ and $e_h$.

During the radiation era, the relation between the coefficient $C_1$ and 
the gravitational potential is simply, using once more (\ref{25a}), 
\beq
\Phi={2\over 3} C_1(\k).
\eeq
Therefore, the gravitational potential during the radiation era 
for modes larger than the Hubble scale is given by the expression \cite{ps92}
\beq
\hat\Phi\simeq -{4\sqrt{\pi G}\over 3}k^{-3/2}\sqrt{s_k}
H_k\left[\sin\theta_k e_h(\k)+\cos\theta_k
e_l(\k)\right],
\label{27}
\eeq
where $H_k$ is given as a function of $s_k$ in (\ref{6}) and 
$s_k$ is given as a function of $k$ in (\ref{17}).
The hat in the above equations (and in all subsequent equations)
indicates that the value of the corresponding quantity is taken deep
in the radiation era when the wavelength of the Fourier mode
 is larger than the Hubble radius.

\subsection{Isocurvature initial perturbations}
As explained in subsection A, the isocurvature perturbations in the present 
 model are due to variations in the relative proportions 
of cold dark matter, generated by the heavy scalar field, with respect to  the
 three other main  species (photons, baryons, neutrinos), 
all generated by the light scalar field.
Moreover, during the radiation era, the isocurvature perturbation
$S$ rigorously defined by (\ref{23}) is essentially the comoving 
cold dark matter density contrast, $S\simeq\dc_c$, so that what is needed
to obtain the primordial isocurvature spectrum is  simply to 
compute the cold dark matter density contrast in terms of the 
scalar field perturbations during inflation. This task was carried out  
in \cite{ps94}. Only the main points will be summarized here.

Let us first give the comoving energy density  perturbation 
associated with the heavy scalar field:
\beq
\d\rho_h^{(c)}=\dot\phi_h\d\dot\phi_h+m_h^2\phi_h\d\phi_h+
3H\dot\phi_h\d\phi_h-\dot\phi_h^2\Phi. \label{28}
\eeq
Matching the inflationary phase when $\phi_h$ is slow-rolling to the 
the inflationary phase when $\phi_h$ is oscillating and then to the 
post-reheating radiation dominated phase, one finds \cite{ps94}
\beq
\d_h^{(c)}\simeq -{4\over 3}m_h^2C_3. \label{29}
\eeq

The coefficient $C_3$ can then be obtained, during inflation, by substracting
the two equations in (\ref{15}) and then using the (slow-roll) background
equations of motion and (\ref{16}).
Inserting the result in (\ref{29}), the density contrast of 
the cold dark matter 
(associated with the heavy scalar field), for modes larger 
than the Hubble radius,  is found to be  given, during the 
radiation era,  by the expression
\beq
\d_h^{(c)}\simeq \sqrt{2\over k^3}H_k\left(\phi_h^{-1} e_h(\k)-{m_h^2\over m_l^2}
\phi_l^{-1} e_l(\k)\right),
\eeq
where the value of the scalar fields is taken at Hubble radius crossing.
This can be reexpressed, using (\ref{3}) and (\ref{5}),  in the form
\beq
\hat S\simeq \d_h^{(c)}\simeq 
2\sqrt{\pi G}k^{-3/2}s_k^{-1/2}H_k\left[{e_h\over \sin\theta_k}
-{R^2 \over \cos\theta_k}e_l\right].
\label{31}
\eeq
Note that the isocurvature perturbations have the same power-law dependence 
as the adiabatic perturbations multiplied by  a weakly k-dependent  
expression which is different from the analogous  expression in (\ref{27}).

\subsection{Conditions for the existence of correlated adiabatic and 
isocurvature perturbations}
As shown above, the quantities describing the primordial adiabatic 
and isocurvature perturbations are in general linear combinations 
of the independent stochastic quantities $e_l$ and $e_h$ and are thus expected 
to be correlated. It is now 
necessary to examine the actual value of the corresponding coefficients.
For adiabatic perturbations, i.e. in equation (\ref{27}), 
the light contribution is dominant for 
$\tan\theta<1$ whereas the heavy contribution is dominant for 
$\tan\theta>1$.  For isocurvature perturbations, i.e. in equation (\ref{31}),
 the light contribution 
dominates for $\tan\theta>R^2$ whereas the heavy contribution is 
predominant in the opposite case. Assuming $R^2\gg 1$, one can thus divide the 
space of parameters for double inflation into three regions:

\subsubsection{Region $\tan\theta>> 1$}

The adiabatic perturbations are dominated by the heavy scalar field 
while the isocurvature perturbations are dominated by the light scalar 
field. The two types of perturbations will thus appear independent.
Moreover, except for $\theta$ very close to $\pi/2$, the isocurvature 
amplitude will be suppressed with respect to the adiabatic amplitude 
by a factor $s_k$. In this parameter region, one recovers the standard 
results of a pure adiabatic spectrum due to a single scalar field, here
the heavy scalar field.

\subsubsection{Region $\tan\theta<< R^{-2}$}

In this region, essentially only the light scalar field contributes to 
the adiabatic perturbations while the isocurvature perturbations are 
dominated by the heavy scalar field. The two contributions are therefore
independent and the isocurvature amplitude can be very high with 
respect to the adiabatic one if $\theta$ is sufficiently small.

\subsubsection{Region $R^{-2}\leq \tan\theta \leq  1$}
This is the most interesting region. Here, both the adiabatic and 
isocurvature perturbations are essentially feeded by the fluctuations
of the light scalar field (even if their amplitude depends on the 
background value of the two scalar fields). This means that the 
adiabatic and isocurvature perturbations are strongly correlated in 
this region. If one considers  the relative magnitude of these 
light scalar field contributions, 
one sees that the isocurvature contribution can compensate 
the $s_k^{-1}$ suppression (with respect to the adiabatic perturbations) 
by a  suitable factor $R^2$. 
Note also that in the upper part of this parameter region, 
i.e. for $\tan\theta \lesssim 1 $,
the heavy and light contributions in the adiabatic perturbations will be 
of similar order while the heavy contribution in the isocurvature 
perturbations  can be ignored. In contrast, in the lower 
part of the region, i.e. $\theta\sim R^{-2}$, 
the heavy contribution in the adiabatic perturbations is negligible whereas
the light and heavy contributions in the isocurvature perturbations are 
of similar weight. This is illustrated on  Figure 1, which displays
 the relative behaviour of the four contributions 
as a function of the angle $\theta$.

The expressions for the adiabatic and isocurvature primordial perturbations,
(\ref{27}) and (\ref{31})  
can be simplified further when one assumes  that these perturbations
 are produced during 
some  specific phases 
of inflation. For instance, if all scales of cosmological 
relevance are produced during the period  of inflation dominated by the heavy 
scalar field, then the approximate relation (\ref{5bis}) relating the angle $\theta$
to the number of e-folds applies, which enables us to simplify the 
Hubble parameter expression, given in equation (\ref{6}),  into
\beq
H(s)\simeq \sqrt{2\over 3}m_l\sqrt{R^2-1}\sqrt{s-s_0}.
\eeq
The various contributions to adiabatic and  isocurvature perturbations then 
reduce to the form 
\beq
k^{3/2}\hat\Phi_h\simeq -{4\sqrt{6\pi G}\over 9}m_l\sqrt{R^2-1}\left(s_k-s_0
\right), \quad k^{3/2}\hat\Phi_l\simeq -{4\sqrt{6\pi G}\over 9}m_l
\sqrt{R^2-1}\sqrt{s_0\left(s_k-s_0
\right)},
\eeq
and 
\beq
k^{3/2}\hat S_h\simeq {2\sqrt{6\pi G}\over 3}m_l\sqrt{R^2-1},
\quad
k^{3/2}\hat S_l\simeq -{2\sqrt{6\pi G}\over 3}m_lR^2\sqrt{R^2-1}
\sqrt{s_k-s_0\over s_0},
\eeq
where the indices $h$ and $l$ refer to the corresponding coefficients 
of $e_h$ and $e_l$ in (\ref{27}) and (\ref{31}).
Let us briefly comment these results when one varies 
the free parameters of the model, $m_l$, $R$ and $s_0$ (but remaining 
in the domain of validity of the above approximate expressions).  
Considering  the variations with respect to the first two parameters, one can 
notice that all the contributions
are proportional to the term $m_l\sqrt{R^2-1}$, except $\hat S_l$ which 
contains an additional $R^2$ dependence. This means, ignoring for the moment
the (weak) scale dependence, that the relative amplitudes 
 of three of the contributions are fixed, the relative amplitude of 
$\hat S_l$ being adjustable by the mass ratio $R$. 
Once $R$ is fixed, the overall 
amplitude of the perturbations can be fixed by the scale $m_l$. 
Concerning  now the variation of the contributions with the cosmological 
scale, $\hat S_h$ is scale-invariant, while the three other are weakly scale 
dependent: $\hat \Phi_l$ and $\hat S_l$ have the same dependence, whereas 
$\hat \Phi_h$ has a stronger dependence.

Another limiting case corresponds to $\theta\ll R^{-1}$, which 
occurs during the period of inflation dominated by the light scalar field. 
In this case, one has $s\simeq s_0 \theta^{2/R^2}$ and the Hubble 
parameter is approximately given by
\beq
H(s)\simeq  \sqrt{2\over 3}m_l\sqrt{s}.
\eeq
As a consequence, the 'heavy' and 'light' contributions are approximated by 
\beq
k^{3/2}\hat\Phi_h\simeq -{4\sqrt{6\pi G}\over 9}m_l s_k \theta_k,
\quad 
k^{3/2}\hat\Phi_l\simeq -{4\sqrt{6\pi G}\over 9}m_l s_k
\eeq
for the adiabatic perturbations and
\beq
k^{3/2}\hat S_h\simeq {2\sqrt{6\pi G}\over 3}m_l\theta_k^{-1},
\quad
k^{3/2}\hat S_l\simeq -{2\sqrt{6\pi G}\over 3}m_lR^2
\eeq
for isocurvature perturbations.
The 'heavy' adiabatic contribution is thus negligible and the 
perturbations due to the heavy scalar field are therefore essentially 
isocurvature.  

Note, to conclude this section, that in their work \cite{ps92}, Polarski
and Starobinsky, concentrated their attention on the intermediate case 
where the scales of cosmological relevance just correspond to the 
transition zone from  the heavy scalar field driven inflation to the 
light scalar field driven inflation. As a consequence, their spectrum 
has a stronger variation in $k$ than in the limiting cases considered above.
Here, the emphasis is put on  contributions to the isocurvature 
perturbations. With another choice of parameters, one can also produce
a huge temperature anisotropy dipole due to isocurvature perturbations
on scales larger than the present Hubble radius \cite{l96}.

\section{Spectra and correlation of adiabatic and isocurvature perturbations}

\subsection{General definitions}
It is usually assumed  in cosmology  that the perturbations can be described
by (homogeneous and isotropic) 
 gaussian random fields. In the specific model under consideration here,  where 
the perturbations are created during an inflationary phase, this is true 
by construction. What is new here is that  
isocurvature and adiabatic perturbations are not assumed to be  independent. 
Indeed, as  shown in the 
previous section, in the case of double inflation, the two kinds of 
perturbations are correlated, at least for some region of the parameter space.
It will thus be our purpose to define statistical quantities that can describe
random fields  which are, a priori, correlated.
Let us first recall, for any homogeneous and isotropic random field 
$f$, the standard definition (up to a normalization factor) 
 of its power spectrum by the expression
\begin{equation}
\langle f_{\k}f_{\k'}^*\rangle = 2\pi^2 k^{-3}{\cal P}_f(k)
\delta(\k-\k').\label{42}
\end{equation}
In addition to this definition, it will be useful to define a 
covariance spectrum between two
random fields  $f$ and $g$ by the following expression
\begin{equation}
{\cal R}e\langle f_\k g^*_{\k'}\rangle  = 2\pi^2 k^{-3} {\cal C}_{f,g}(k)
\delta(\k-\k').
\end{equation}
In order to estimate the degree of correlation between two quantities, 
it is convenient to  
also define the correlation spectrum ${\tilde{\cal C}}_{f,g}(k)$  
by normalizing ${\cal C}_{f,g}(k)$:
\beq
{\tilde{\cal C}}_{f,g}(k)={{\cal C}_{f,g}(k)\over \sqrt{{\cal P}_f(k)}
\sqrt{{\cal P}_g(k)}}. \label{44}
\eeq
Schwartz inequality implies, as usual, that 
$-1\leq {\cal C}_{f,g}(k) \leq 1$. 
The correlation (anticorrelation) will be stronger as one is closer to $1$ or 
$-1$.

\subsection{Double inflation generated perturbations}
Let us now specialize the above formulas to the case of perturbations 
generated by double inflation. By substituting the explicit expressions for
the perturbations obtained in the previous section, namely 
(\ref{27}) and (\ref{31}), one finds
\beq
{\cal P}_{\hat\Phi}={8G\over 9\pi}H_k^2s_k
\eeq
for the initial adiabatic spectrum,
\beq
{\cal P}_{\hat S}={2G\over \pi}{H_k^2\over s_k}\left[{R^4\over \cos^2\theta}
+{1\over\sin^2\theta}\right],
\eeq
for the initial isocurvature spectrum and
\beq
{\cal C}_{\hat\Phi,\hat S}={4 G\over 3\pi}H_k^2(R^2-1)
\eeq
for the covariance spectrum. Combining the three above spectra 
according to (\ref{44}), one finds finally 
for the correlation spectrum the  expression
\beq
{\tilde {\cal C}}_{\hat\Phi,\hat S}={(R^2-1)\sin{2\theta}\over 2(R^4
\sin^2\theta+\cos^2\theta)^{1/2}}.
\eeq
It is instructive to study the dependence of this correlation spectrum 
with respect to the parameters of the model. If one takes $\theta$ fixed, 
one sees that the correlation will vanish for $R=1$ and will then 
increase monotonously with increasing $R$ approaching the asymptotic 
value $\cos\theta$.
If now one considers $R$ as fixed and study the variations of the 
correlation with respect to  $\theta$, one 
 recovers the conclusions of section 3D: the correlation 
vanishes when $\theta$ approaches zero or  $\pi/2$; inbetween, one can 
see that the correlation reach a maximum  for $\sin^2\theta=(R^2+1)^{-1}$,
with  the  value  
\beq
{\tilde {\cal C}}_{\hat\Phi,\hat S}^{max}={R^2-1\over R^2+1}.
\eeq
The correlation spectrum ${\tilde {\cal C}}_{\hat\Phi,\hat S}$ for 
various choices of 
parameters has been plotted on Fig. 2, as a function of $s_k$. One can, 
as before, distinguish between the two extreme cases. For models such 
that $\theta\gg R^{-1}$, corresponding to a 'heavy' inflationary phase, the 
various contributions vary slowly with $\theta$, as can be seen from 
Fig. 1. This means that the correlation will be almost constant.
For models such that $\theta \ll R^{-1}$, corresponding to a 'light' 
inflationary phase, $\hat S_h$ increases quickly with decreasing $\theta$, i.e. 
with decreasing scales, which implies that the correlation will decrease 
with decreasing scales, i.e. smaller $s_k$. The models with $s_0< s_H$ 
belong to the first category, while models with $s_0> s_H$ correspond
to the second. Finally, the models with $s_0$ close to $s_H$ have 
an intermediate behaviour between the two extreme cases. They also have 
the strongest correlation.

\section{Predictions for the CMBR and density contrast spectrum}
\subsection{Analytical predictions for long-wavelength perturbations}
\subsubsection{Evolution of the perturbations}
In the case of perturbations whose wavelength  is 
 larger than the Hubble radius, the time  evolution is particularly 
simple. For an initial isocurvature 
perturbation characterized by the initial amplitude $\hat S$,
 the entropy perturbation $S$ is unchanged as long as the 
perturbation is larger than the Hubble radius, whatever the evolution 
of the backgroung equation of state, i.e.
\beq
S=\hat S \qquad (k\ll aH). \label{51}
\eeq
 However, the radiation-matter 
transition will generate a gravitational potential perturbation (see e.g. 
\cite{ll})
\begin{equation}
\Phi^{iso}=-{1\over 5}\hat S\qquad (k\ll aH). \label{52}
\end{equation}
Of course, the initial adiabatic perturbation 
will also contribute to the gravitational 
potential perturbation:
\beq 
\Phi^{ad}=T\hat \Phi \qquad (k\ll aH),\label{53}
\eeq
where $T$ is a coefficient, close to $1$, due to the evolution of the 
universe (if one ignores the anisotropic stress of the neutrinos, 
$T=9/10$.)

\subsubsection{Large angular scale CMBR anisotropies.}
At large angular scales, the temperature anisotropies are essentially
due to the sum of an intrinsic contribution and of a Sachs-Wolfe \cite{sw}
contribution.
Except for the dipole for which the Doppler terms are important, 
the Sachs-Wolfe contribution can be written (for a spatially
flat background)
\begin{equation}
\left(\Delta T\over T \right)_{SW}(\e)={1\over 3}\Phi(x_{ls});
\end{equation}

Where $\e$ on the left hand side is a unit vector corresponding
to the direction of observation and $x_{ls}$ on the right hand side
represents the intersection of the last scattering surface with the light-ray
of direction $\e$.
The intrinsic contribution is simply given, via the Stefan law, as the
perturbation $\dc_\gamma/4$ at the time of last scattering. Since last
scattering occured in the matter era, $\dc_m \simeq \dc$,
and therefore for an adiabatic perturbation 
($\dc_m = {3\over 4}\dc_\gamma$), $\left({\Delta T}\over
T\right)_{\rm int} \simeq {1\over 3} \dc_m$, which can be seen to be
negligible (see below (\ref{poisson})) with respect to the
Sachs-Wolfe contribution, whereas for an isocurvature perturbation
($S\simeq -{3\over 4}\dc_\gamma$ during matter era), 
\begin{equation}
\left(\Delta T\over T\right)_{int}\simeq -{1\over 3} S.
\end{equation}
To conclude, the  temperature anisotropies will be given in
general by
\begin{equation}
{\Delta T\over T}={1\over 3}T\hat\Phi - {2\over 5}\hat S \label{56}
\end{equation}
on angular scales larger than the angle (of the order of the degree) 
corresponding to the size of the Hubble radius at the time of the last 
scattering.
This equation enables us to estimate 
easily the normalization of the temperatures anisotropies for the low
multipoles (see the definition (\ref{62})-(\ref{63})), essentially 
constrained by COBE measurements.
Note that for mixed primordial perturbations with isocurvature and 
adiabatic contributions of the same order of magnitude, the 
 low multipoles anisotropies can be significantly reduced by a compensation 
effect between the isocurvature perturbation and the adiabatic one. 
It turns out that  this is the case for the light scalar field contribution 
in double inflation models with 
$R\sim 5$ (see Fig.1 and the consequence on Fig. 3-5). 

\subsubsection{Large scale structure}

Large scale structure is governed by the density contrast, or equivalently
the gravitational 
 potential perturbation $\Phi$ since the latter quantity 
can be related to the (total) contrast
density in the comoving gauge by the (generalized) Poisson equation, which 
reads (see e.g. \cite{mfb})
\beq
\left({k\over a H}\right)^2\Phi=-{3\over 2}\dc. \label{poisson}
\eeq 
 For modes inside the Hubble radius
the evolution becomes quite complicated and depends on the specific
ingredients of the model. But what is relevant for our purpose is that this
subhorizon evolution does not depend on the nature of the primordial
perturbations. What matters is the total gravitational potential
perturbation $\Phi$ which can be written, in the matter era, as
\begin{equation}
\Phi=\Phi_{ad}-{1\over 5}S. \label{57}
\end{equation}
Note  that the influence of primordial isocurvature perturbations is 
smaller on the large scale density power spectrum  (see (\ref{57})) than 
on large scale temperature anisotropies (see (\ref{56})).

\subsubsection{Spectra}

Using the relation (\ref{56}) (with $T=1$), the spectrum for the large 
scale temperature
anisotropies can be expressed in terms of the primordial isocurvature 
and adiabatic spectra, 
\begin{equation}
{\cal P}_{{\Delta T\over T}}={1\over 9}{\cal P}_{\hat\Phi}
+{4\over 25}{\cal P}_{\hat S} -{4\over 15} {\cal C}_{\hat\Phi,\hat S}.
\end{equation}
When only primordial adiabatic perturbations are present, the previous
expression implies
\begin{equation}
{\cal P}^{1/2}_{\Delta T\over T}={1\over 3}{\cal P}^{1/2}_\Phi ,
\end{equation}
whereas for pure isocurvature perturbations, one finds
\begin{equation}
{\cal P}^{1/2}_{\Delta T\over T}=2{\cal P}^{1/2}_{\Phi} .
\end{equation}
This is in agreement with  the standard comment in the literature that isocurvature
perturbations generate CMBR anisotropies six times bigger than equivalent
adiabatic perturbations.
This is the reason why isocurvature perturbations are in general rejected
in comological models \cite{eb86}. However, when one takes into account both
isocurvature and adiabatic perturbations, with the possibility of
correlation, then the additional term due to correlation can change 
significantly these conclusions. Illustrations will be given in the next 
subsection.

Similarly, the spectrum for the gravitational
potential is given by
\begin{equation}
{\cal P}_{\Phi} = {\cal P}_{\hat\Phi}+{1\over 25}{\cal P}_{\hat S}-{2\over 5}
{\cal C}_{\hat\Phi,\hat S} .
\end{equation}

\subsection{All-scale predictions}
After having considered long wavelength  perturbations, whose advantage 
is one can  estimate analytically their observable amplitude and 
thus normalize easily  the models, let 
us analyze now smaller scales, which require the use of numerical computation.

\subsubsection{CMBR anisotropies}
As it is customary, one decomposes the CMBR anisotropies on the basis 
of spherical harmonics:
\begin{equation} 
{\Delta T\over T}(\theta, \phi)=
\sum_{l=1}^{\infty}\sum_{m=-l}^{l} a_{lm}Y_{lm}(\theta,\phi). \label{62}
\end{equation}
The predictions of a model are usually given in terms of the expectation 
values of the squared multipole coefficients
\beq
C_l\equiv \langle |a_{lm}|^2\rangle. \label{63}
\eeq 
In the present model, the temperature anisotropies will be the superposition 
of a contribution due to the heavy scalar field and of a contribution 
due to the light scalar field. These two contributions are independent, because
the stochastic quantities $e_h$ and $e_l$ are independent,  and
as such the coefficients $C_l$ can be decomposed 
\beq
C_l=C_l^{(l)} +C_l^{(h)},
\eeq
where the upper index refers to the 'light' or 'heavy' nature of the 
perturbations. It is important to emphasize that only a decomposition of 
this type is allowed here. For example, a decomposition of the $C_l$ as a 
sum of an isocurvature contribution and of an adiabatic contribution 
would be wrong here.
In practice, 
the heavy and 
light contributions to the $C_l$ are computed independently, by using twice
a Boltzmann code (developped in our group by A. Riazuelo, and used in 
\cite{udr98}). 
The first run takes as initial condition $\hat \Phi_h$ 
and $\hat S_h$ and yields
the coefficients $C_l^{(h)}$. Similarly, the second run computes the 
$C_l^{(l)}$ using as initial conditions the corresponding quantities 
$\hat \Phi_l$ and $\hat S_l$.
The results for $C_l^{(l)}$ and $C_l^{(h)}$, as well as their 
sum $C_l$, are plotted on Fig. 3-6 for four illustrative models (for all models
the Hubble parameter and the baryon density correspond 
respectively to  $h_{100}=0.5$, $\Omega_b=0.052$). 
For the first three models, the value $R=5$ has been chosen
 because the isocurvature 
and adiabatic contributions of the light scalar field are then of similar 
amplitude, as is visible on Fig. 1, and the effects of mixing and correlation 
are particularly important. A consequence of the similar amplitude (with the 
same sign) 
of the two `light' 
contributions is an important suppression of the light spectrum  $C_l^{(l)}$ for 
small $l$, as noticed already in the previous subsection, and as is visible 
on Fig 3-5. In contrast, one can check on Fig. 6 that this will not be the case
for the $R=10$ model, for which  $\hat S_l$ is dominant. 

The first two graphs have a roughly similar behaviour for the 
'heavy' contribution. What distinguishes them is the 'light' contribution, 
which illustrates its high sensitivity on the relative amplitudes of 
$\hat S_l$ and $\hat \Phi_l$. A systematic investigation of the effects of 
mixed correlated primordial spectra, independently of the early universe 
model to produce them, on the temperature anisotropies will be given elsewhere
\cite{lr99}. For these two models, one notices an amplification (weak in the 
first case and strong in the second) of the main 
acoustic peak with respect to the standard (pure adiabatic and scale-invariant)
model. In contrast, the third example shows a suppression of the main peak, 
which is due to a strong contribution $\hat S_h$, which makes the 'heavy'
spectrum look ``isocurvature" and thus damps the main peak in the global 
spectrum. Finally, the last example is characteristic of the domination 
of the `light' spectrum, itself dominated by the isocurvature contribution 
($\hat S_l$), which thus makes the global spectrum look "isocurvature".
It is rather remarkable that modest variations of the two relevant parameters of 
the model, $R$ and $s_0$ ($m_l$ is useful simply for an overall normalization 
of the parameters), can lead to a large variety of temperature anisotropy 
spectra.

\subsection{Power Spectrum}
Another quantity which is extremely important for the confrontation of models
with the observations is the  (total) density power spectrum. In the literature, 
it is usually denoted $P(k)$ and its relation to the corresponding spectrum 
${\cal P}_{\dc}$ (for the comoving density constrast) defined  generically 
in (\ref{42}) is 
\beq
P(k)=2\pi^2k^{-3}{\cal P}_{\dc}.
\eeq
Using the Poisson equation (\ref{poisson}), it can be reexpressed in terms of 
the gravitational potential spectrum
\beq
P(k)={8\pi^2\over 9(a_0H_0)^4}k {\cal P}_\Phi(k).
\eeq

As with the temperature anisotropies, the power spectrum for double inflation
 is obtainable by computing independently the power 
spectrum for the heavy scalar field contribution, then  that for the light 
scalar field contribution, and finally by adding the two results,
\beq
P(k)=P_l(k) + P_h(k).
\eeq
In contrast with the temperature anisotropies, the influence of the 
mixing and correlation of the primordial perturbations on the density 
spectrum is less spectacular because the shapes of 
the pure isocurvature and pure adiabatic density spectra 
are not extremely different. There is however a sensible difference:
the pure isocurvature spectrum has, relatively to the large scales, less
power on small scales than the pure adiabatic spectrum. To illustrate 
what happens for mixed and correlated primordial perturbations, Fig. 7 
displays, for  the  model specified by the parameters $R=5$ and $s_0=50$, 
the total power spectrum , together with the two independent `light' and `heavy'
contributions, as well as  the standard adiabatic CDM 
power spectrum (normalized as before)
for comparison. Note that the resulting spectrum has, relatively 
to large scales, less power than the standard adiabatic power spectrum. 
However, it has globally more power than the standard spectrum for the 
same temperature anisotropy amplitude on small $l$.

\section{Conclusions}
The main conclusion of this work is that it is possible, in the simplest 
model of multiple inflation, to obtain correlated isocurvature and 
adiabatic primordial perturbations. These perturbations slightly 
deviate from scale-invariance but their correlation can entail 
significant modifications with respect to standard single scalar field
models. 

This class of models, both simple and rich, could provide an 
interesting field of experiment to investigate the feasibility 
to determine the cosmological parameters and the primordial perturbations 
from expected data. The question would
be, assuming Nature has chosen this particular model, could we infer 
from the expected temperature anisotropy data  the cosmological parameters
and to which precision?  More important, would it possible to discriminate 
between a single field inflation model and a multiple field model 
with correlated perturbations and what would be the price to pay on the 
precision of the cosmological parameters ?

It was not the purpose of the present work to exhibit a model supposed 
to fit better the observations. However, one of these models, surprisingly,
turns out to present two characteristics, which are at present favoured 
by observations: a power spectrum with modest power at small scales 
(comparatively to the standard CDM model) and a high peak on intermediate 
scales. It may be worth seeing how well this model does when confronted 
with the current observations.

Finally, it would be interesting to investigate the possiblity of 
correlated adiabatic and isocurvature perturbations within the framework 
of multiple inflation with interaction between scalar fields and see 
how the main features presented here would be modified.

\acknowledgements
I would like to thank Alain Riazuelo for his help with the Botzmann code he has 
developed.

\begin{figure}
\centering
\epsfig{figure=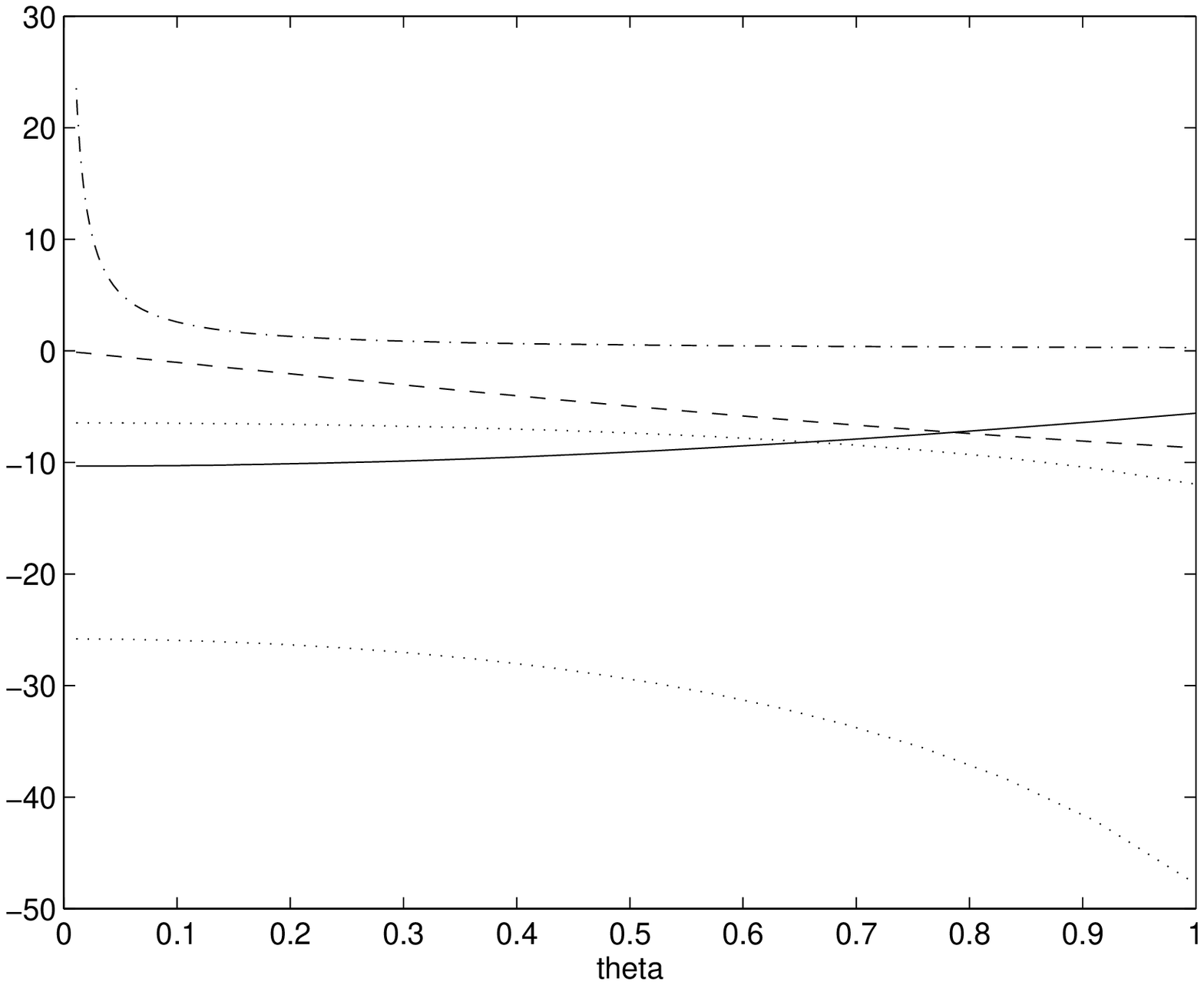, width=9cm}
\caption{Relative amplitude of the $\phi_h$ contributions to the adiabatic
perturbations, $\hat \Phi_h$ (dashed line), to the isocurvature perturbations,
 $\hat S_h$ (dotted 
dashed line), and of the $\phi_l$ contribution to the adiabatic perturbations, 
$\hat \Phi_l$
(continuous line), and to the isocurvature perturbations, $\hat S_l$
 (dotted line). 
The last quantity, the only one which depends on $R$, has been plotted for 
$R=5$ (upper dotted line) and $R=10$ (lower dotted line).}
\end{figure}

\begin{figure}
\centering
\epsfig{figure=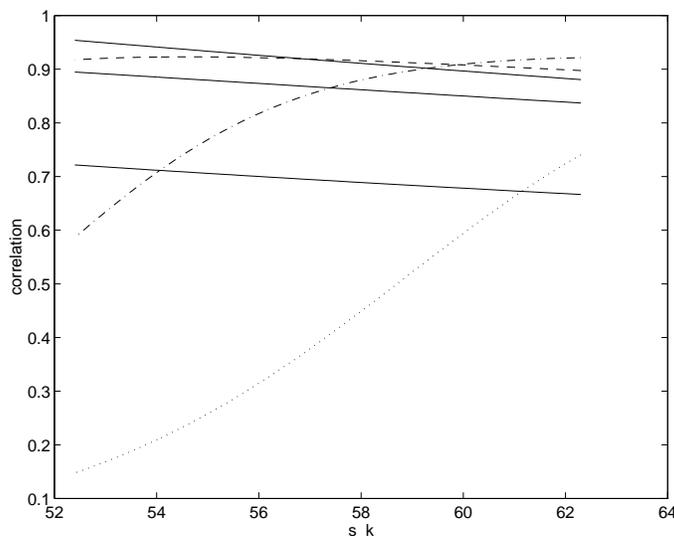, width=9cm}
\caption{Correlation spectrum for various parameters. Continuous curves 
from bottom to top (on the left hand side of the figure) correspond respectively 
to $(R=5,s_0=30)$, $(R=5,s_0=50)$ and $(R=10,s_0=50)$. The dashed curve 
corresponds to $(R=5,s_0=60)$, the dotted dashed curve to $(R=5,s_0=70)$ 
and finally the dotted curve to $(R=5,s_0=80)$.}
\end{figure}

\begin{figure}
\centering
\epsfig{figure=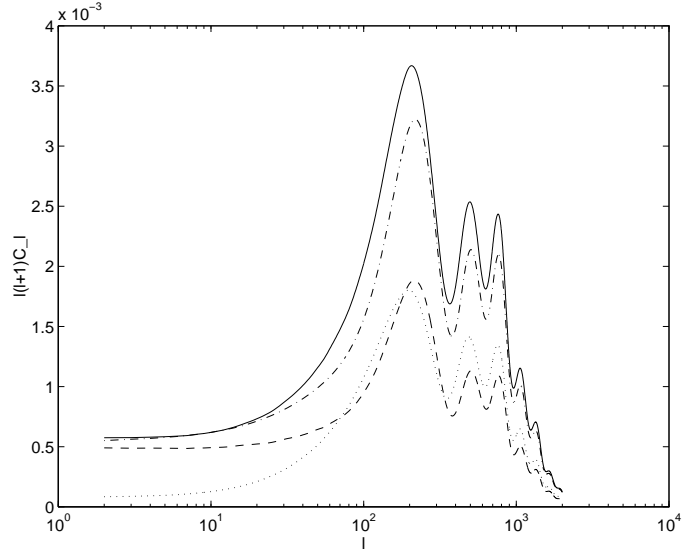, width=9cm}
\caption{Temperature anisotropies for the double inflation scenario with 
$R=5$, $s_0=30$. The total anisotropies (continuous line) are the sum
of a contribution due to the heavy scalar field (dashed line) and of a 
contribution of the light scalar field (dotted line). To make the comparison, 
the anisotropies due to standard (adiabatic scale-invariant) perturbations
are also plotted (dotted dashed line), using  $C_{10}$ for normalization.}
\end{figure}
\begin{figure}
\centering
\epsfig{figure=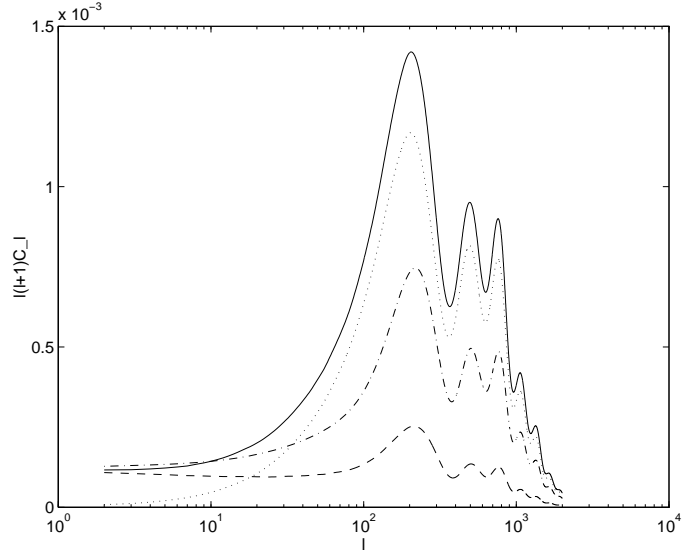, width=9cm}
\caption{Temperature anisotropies for the double inflation scenario with 
$R=5$, $s_0=50$ (same conventions as in Fig.3).}
\end{figure}
\begin{figure}
\centering
\epsfig{figure=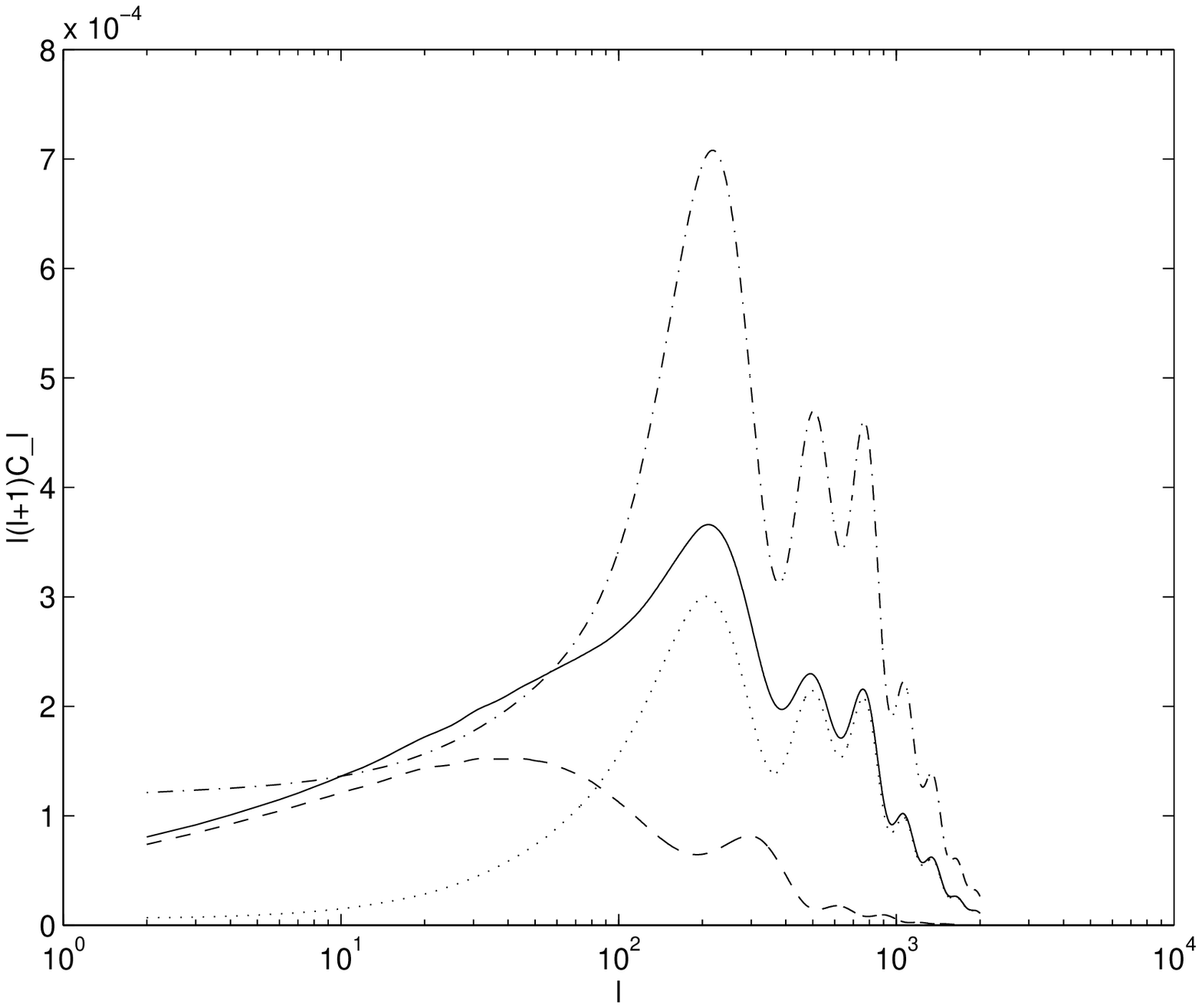, width=9cm}
\caption{Temperature anisotropies for the double inflation scenario with 
$R=5$, $s_0=80$.}
\end{figure}
\begin{figure}
\centering
\epsfig{figure=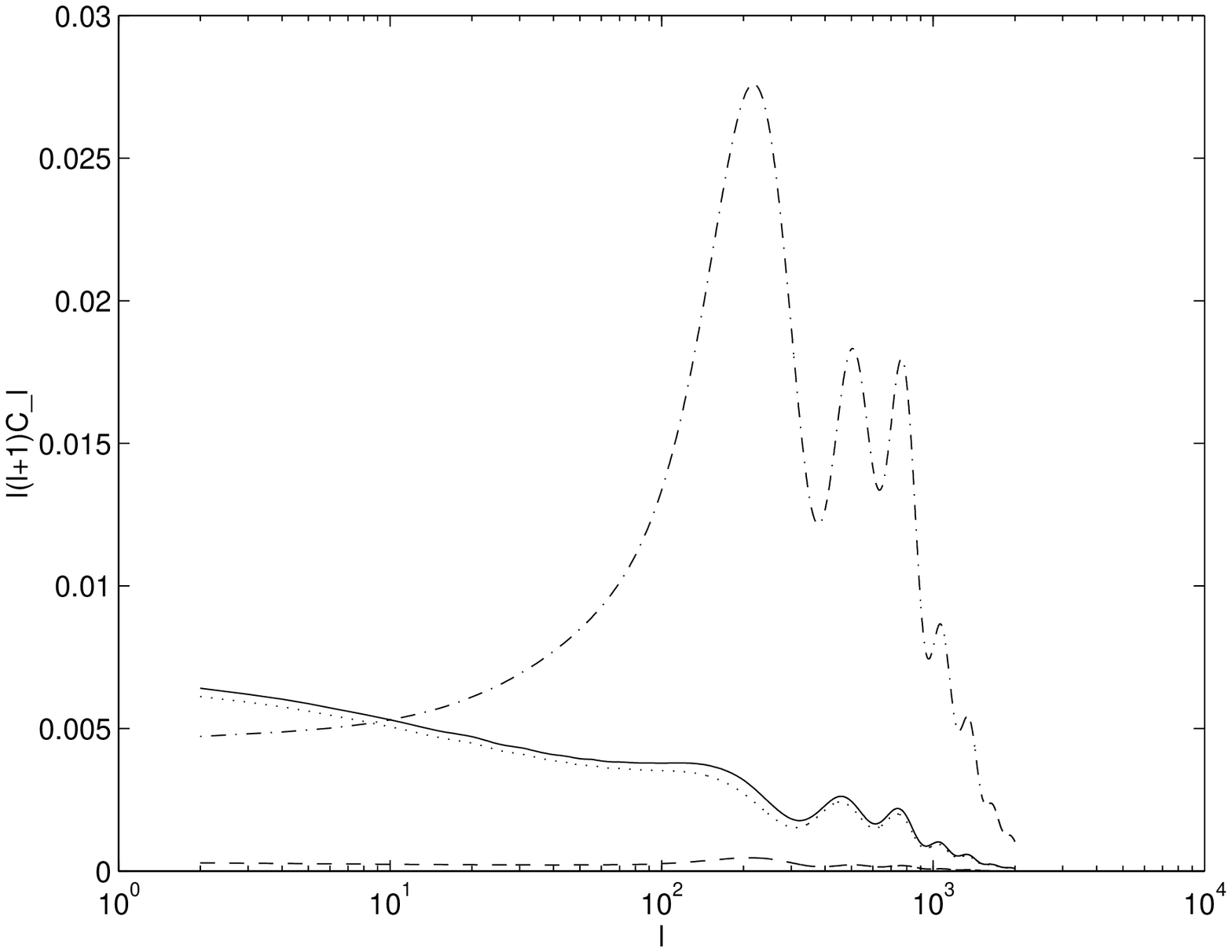, width=9cm}
\caption{Temperature anisotropies for the double inflation scenario with 
$R=10$, $s_0=50$.}
\end{figure}

\begin{figure}
\centering
\epsfig{figure=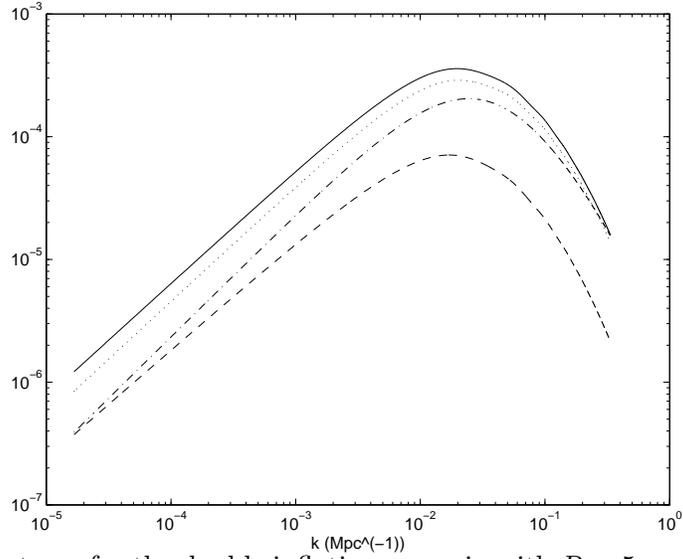, width=9cm}
\caption{Power spectrum  for the double inflation scenario with 
$R=5$, $s_0=50$. The total power spectrum (continuous line) is the sum
of a contribution due to the heavy scalar field (dashed line) and of a 
contribution of the light scalar field (dotted line). The standard 
(adiabatic scale-invariant) power spectrum is  also plotted 
(dotted dashed line) with the same normalization for $C_{10}$.}
\end{figure}

\end{document}